# Optimal particle type and projection number in multi-modality relative stopping power acquisition

Marta F. Dias[1;2;a], Charles-Antoine Collins-Fekete[2;3;4], Lennart Volz[2;5], Guido Baroni[1;6], Marco Riboldi[1], Joao Seco[2;5]

1. Dipartimento di Elettronica, Informazione e Bioingegneria - DEIB, Politecnico di Milano, Italy
2. Deutsches Krebsforschungszentrum, Heidelberg, Baden-Wurttemberg, Germany
3. Departement de physique, de genie physique et d'optique et Centre de recherche sur le cancer, Universit Laval, Quebec, Canada
4. Departement de radio-oncologie et CRCHU de Quebec, CHU de Quebec, QC, Canada
5. University of Heidelberg, Department of Physics and Astronomy Heidelberg, Baden-Wurttemberg, Germany
6. Bioengineering Unit, Centro Nazionale di Adroterapia Oncologica, Pavia 27100, Italy
a. Corresponding author: marta.dias@polimi.it

**Abstract**

Proton radiography combined with X-ray computed tomography (CT) has been proposed to obtain a patient-specific calibration curve and reduce range uncertainties in cancer treatment with charged particles. The main aim of this study was to identify the optimal charged particle for the generation of the radiographies and to determine the optimal number of projections necessary to obtain minimal range errors. Three charged particles were considered to generate the radiographies: proton (p-rad), helium ions (α-rad) and carbon ions (c-rad). The problem formulation was viewed as a least square optimization, $argmin_x \left( \left\| A\vec{x} - \vec{b} \right\|_2^2 \right)$, where the projection matrix $A$ represents the cumulative path crossed by each particle in each material, $\vec{b}$ is the charged particle list-mode radiography expressed as the crossed water equivalent thickness (WET) and $\vec{x}$ is the relative stopping power (RSP) map to be optimized. The charged particle radiographies were simulated using the Geant4 Monte Carlo (MC) toolkit and an anthropomorphic adult head phantom. $A$ was determined by calculating the particles trajectory through the X-ray CT employing a convex hull detection combined with a cubic spline path. For all particle types, the impact of using multiple projections was assessed (from 1 to 6 projections). In this study, tissue segmentation methods were also investigated in terms of achievable accuracy. The best results (mean RSP error below 0.7% and range errors below 1 mm) were obtained for a-rad when 3 projections were used. The dose delivered to the phantom by a single α-rad was 8 µGy, lower than the dose of X-ray radiography.

# 1. Introduction

The finite range of charged particles, like protons, helium and carbon ions, represents one of the main reasons for their usage in cancer therapy. With precise range calculation, it is theoretically possible to deliver a high and conformal dose to the tumor volume while sparing organs at risks (OARs) [1]. However, due to uncertainties affecting the range estimate inside the patient, the full potential of charged particle radiotherapy cannot be fully exploited [2].

One of the uncertainties stems from the fact that the beam range prediction is based on a conversion of X-ray computed tomography (CT) Hounsfield Units (HUs) to relative stopping power (RSP) [3]. For each patient, the range of the charged particle beam is then calculated using the HU-RSP calibration curve.

The commonly used method to determine the HU-RSP curve is the stoichiometric approach [4]. This approach uses experimentally obtained RSP values for plastic materials to determine a theoretical HU-RSP bijection for human biological tissues. The method neglects inter-patient tissue variation such as: differences in age, sex, diet, or health state which may very likely result in variability of the chemical composition of tissues [5]. Therefore, the stoichiometric calibration is not patient specific and range prediction errors are expected. Yang et al. [5], showed that RSP uncertainties due to this calibration procedure can be as large as 5%.

Recently, Collins-Fekete *et al.* [6] proposed a method which combines a proton radiography (p-rad) with the treatment planning X-ray CT to obtain a patient-specific c calibration curve. In their work, they considered cubic spline trajectories to retrieve the most likely location of protons throughout the patient. By knowing the protons whereabouts, the projected proton energy loss through the X-ray CT can be compared with the energy loss measured in the proton radiography and the RSP of each X-ray CT voxel computed. Their results demonstrated serious potential to increase the accuracy of the clinical HU-RSP calibration.

The purpose of this work was to generalize the method proposed by Collins-Fekete *et al.* [6] to heavier ions, such as carbon and helium. In addition, strategies for selecting a set of representative materials out of the composition of the phantom to be used in the optimization algorithm were studied. Finally, the impact of using multiple angle radiographies in the optimization procedure was assessed.

# 2. Materials and Methods

## 2.1. Determination of the patient specific calibration curve

The method described in Collins-Fekete *et al.* [6] is here shortly summarized. For further details, the reader is referred to the original publication and mentioned references.

Precise knowledge of each particle's path through the patient is paramount to the presented technique. By tracking each particle along its path through the X-ray CT, the information about the total length crossed in each voxel is retrieved and consequently, using the HU-RSP calibration curve, the total WET crossed. This calculated WET can then be compared to the WET measured by the particle radiography. By changing the HU-RSP calibration curve, this difference can be minimized. The HU-RSP optimization problem can be formulated as a linear least squares problem:

$$argmin_x \left(\left\|A\vec{x} - \vec{b}\right\|_2^2\right) \qquad s.t. \qquad x \geq 0. \tag{1}$$

An X-ray CT image is defined by a set of *N* HU numbers whereas, a list-mode particle radiography contains *M* events/trajectories. The system matrix **A** is a *M x N* matrix which contains the $n^{th}$ HU. The vector **x** is a *N*-dimensional vector which represents the RSP value of each HU. The vector **b** is the list-mode projection data (p-rad, α-rad or c-rad) which represents the total WET crossed by the $m^{th}$ particle. $||.||_2^2$ is the squared $\ell^2$ norm. The WET values of the projection vector are obtained by solving the integral over the initial ($E_{init}$) and final energy ($E_{out}$) of each projectile particle:

$$WET = \int_{E_{out}}^{E_{init}} \frac{dE}{SP(I_w, E)}, \tag{2}$$

where $SP(I_w, E)$ is the stopping power of water with Ionization potential $I_w$ and for a particle with kinetic energy $E$. By solving Equation 1 a patient-specific mapping of RSP to HU can be obtained. To solve this optimization problem, the linear least square function of *Python's numpy* package was used [7]. The performance of the optimization was assessed by computing the relative error $\left(100 \times \frac{RSP_{opt} - RSP_{ref}}{RSP_{ref}}\right)$ each obtained RSP value ($RSP_{opt}$) compared to the gold standard reference ($RSP_{ref}$).

The particles' trajectories were estimated using a hull-detection algorithm combined with the optimized cubic spline path estimate proposed by Collins-Fekete et al. [8, 9].

### 2.2. Monte Carlo simulations

Radiographies for proton (p-rad), helium (α-rad) and carbon (c-rad) ions were generated using the Geant4 MC simulation toolkit [10, 11, 12] (v.4.10.2.p01). Standard processes included energy loss and straggling, multiple Coulomb scattering using the Urban model [13] and elastic/inelastic ion interactions from Geant4 ion dedicated packages [14]. Geant4 ion dedicated packages were used, namely the G4HadronElasticProcess, G4HadronInelasticQBBC, G4IonBinaryCascadePhysics, G4IonFluctuations, G4IonParametrisedLossModel and the G4IonQMDPhysics. The implemented method considered the phantom placed in between two ideal detection planes which recorded information about the particles energy, momentum and position before and after crossing the object. No detector effects were considered.

An adult head phantom was used to evaluate the methods presented in this work. The simulated geometry was based on a voxelized X-ray CT geometry acquired of the phantom through the Massachusetts General Hospital (MGH) X-ray CT machine. The tissue composition and density of each voxel of the phantom was extracted following the method described by Collins-Fekete et al. [6]. An initial statistic of $10^6$ primary particles was propagated through the phantom to acquire each radiography. In order to keep the number of secondaries produced by carbon ions crossing the phantom low and as means to guarantee that at least 50% of the incident carbon ions completely crossed the phantom, MC simulations of 450 MeV/u carbon ions were performed. In order not to bias the comparison, the same energy per nucleon was also used for helium and proton particles crossing the adult head phantom. No image reconstruction technique was applied since the method relies only on knowing the particles' whereabouts.

The gold standard, *i.e.* the reference calibration curve, was obtained using the stopping power information calculated directly from the simulated tissue composition. By definition, the RSP of a tissue is given by the stopping power of a particle in that tissue divided by the particle's stopping power in water [15]. For each tissue, the mean ratio was calculated for beam energies between 100 MeV/u to 500 MeV/u. The RSP was approximately constant for the selected energy interval and insensitive to the ion species as shown by Witt and Moyers [16, 17]. Hence, the same reference was used for all investigated charged particles.

### 2.3. Tissue segmentation

The heterogeneity of the human body and the intrinsic CT noise leads to a wide distribution of HU values in the X-ray CT. Optimizing all the HU-RSP values would lead to a large and sparse system matrix (A) which would cause statistical, memory storage and time consuming problems. Therefore, the set of HU values present in the patient X-ray CT needs to be reduced to a smaller sample that still is representative of the complete set. In this work three different approaches were considered to determine a representative set of HU to be optimized.

The first method was based on minimizing the distortion of the HU values sampled from the HU probability density function (PDF). In general, the process of mapping a large set of values into a smaller one is irreversible. This results in loss of information and introduces distortion to the new HU set that cannot be eliminated. However, the distortion can be minimized with the use of a *T*-level quantizer, *i.e.*, the HU sample will be represented by a set of *T* values. The most common distortion measure is given by the mean squared error:

$$D = E[(x - Q(x))^2] = \int_{-\infty}^{\infty}(x - Q(x))^2 f(x)dx = \sum_{k=1}^{T}\int_{-b_k}^{b_k}(x - y_k)^2 f(x)dx, \qquad (3)$$

The quantizer $Q$ is defined by $Q(x) = y_k$, where $y_k$ are the new sampled HU values for the *k*-th interval. $f(x)$ represents the probability density function of the HU in the X-ray CT. $x$ represents the initial complete set of HU. $D$ is the mean squared error over every interval defined by the boundaries $b_k$. To obtain the [$y_k, b_k$] set that minimizes the mean squared error $D$, the Powell minimization algorithm [18] was used with equation 3 as a cost function. More information about distortion and its implementation can be found in [19].

The second method used the *k-means* cluster method from the Python package *sklearn* [20]. The k-means cluster algorithm takes as input the voxelized X-ray CT HUs and a desired number of clusters (*T*), *i.e.*, regions corresponding to a certain tissue composition. The output is a set of *T* cluster centroids which are the HU/RSP values to be optimized. The impact of the number of clusters used in the optimization process was assessed by using *T* = [10; 30; 150] clusters (corresponding to 0.3%, 1% and 5% of the total number of HUs). Finally, the third method implemented considered *T* evenly spaced values out of the PDF from the X-ray CT.

The three applied methods retrieve a set of HU values that were used in the definition of the HU-RSP calibration curves. The HU-RSP pairs not in the calibration curve were extrapolated by linear interpolation. The performance of the different tissue segmentation methods was assessed by comparing the obtained calibration curves to the reference, *i.e.*, gold standard HU-RSP calibration curve.

## 2.4. Range uncertainties

The accuracy in the range calculation with the optimized calibration curves was evaluated using the technique proposed by Bär *et al* [21]. The technique combines Monte Carlo statistical sampling and WET crossed prediction. Heterogeneous material blocks were created combining 1 mm tissue slabs to add up to physical thicknesses between 1 cm and 30 cm. The tissue composition of each slab was sampled from the tissue PDF of the adult head phantom. The WET of each tissue slab was calculated from the optimized calibration curves and compared with the WET calculated using the gold standard reference curve.

## 3. Results

For all results shown, the particle histories were filtered, *i.e.*, particles that were greatly deflected (2 mm displacement from a straight line), were not considered in the optimization process.

### 3.1. The impact of tissue segmentation on optimization accuracy

The X-ray CT of the adult head phantom presents a range of more than 3000 HU values and related RSPs. Three tissue segmentation methods, as described in Section 2.3 were proposed to select a set of representative materials to be optimized and create the HU-RSP calibration curve. The results are presented in Figure 1(a). Each series of results were obtained using three α-rad projections combined with the X-ray CT.

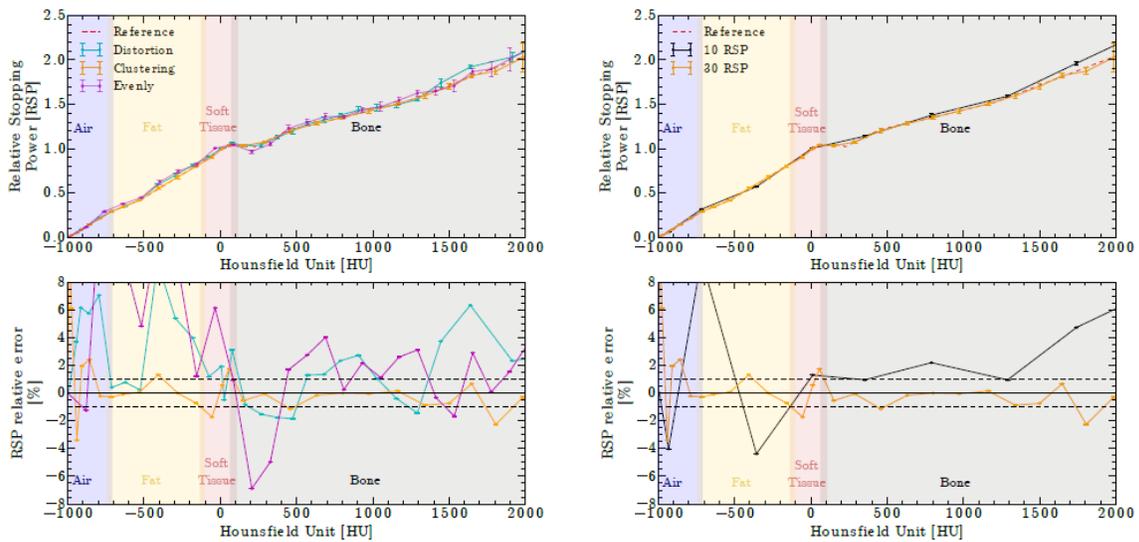

**Figure 1:** Results for the head phantom using α-rad. Figure 1(a) shows the results for the different tissue segmentation methods. Figure 1(b) shows the results using the clustering method with different number of clusters (HU values). For each figure, the topmost graph shows the calculated calibration curves and the MC reference. Bottom: Relative RSP errors between reference (MC data) and the optimized calibration curves.

Sampling the PDF in evenly spaced steps presented the largest errors (mean error of 3.05±2.00%), followed by the distortion method (mean error of 1.55±0.73%). The clustering method showed the smallest relative RSP errors (mean error of 0.62±0.63%)). The mean errors were calculated for HUs between -100 and 1300, which represent more than 90% of the phantom tissues (Figure 1(a)).

Since the best results were obtained using the clustering method, this approach was used to evaluate the impact of the sample size and the results are presented in Figure 1(b). For a set of 10 RSPs, the maximal relative difference (Figure 1(b) bottom) was larger than 4%, whereas the errors were reduced below 1% when using 30 RSPs (for HUs between -100 and 1300). Using the clustering with 150 RSPs to be optimized resulted in very large variations of the calibration curve and, consequently, are not shown in 1(b). Using 150 RSPs led to a mean relative error of 1.09±0.99% compared to 0.62±0.63% for 30 RSP.

### 3.2. The impact of the number of projections/charged particle type on range uncertainties

Given the previous results, only the clustering segmentation method with sample size of 30 RSP values was considered for the subsequent analysis. For each charged particle radiography, the impact of the number of projections (maximum of 6 for angles: 0º, 30º, 45º, 60º, 90º, 130º) as input to the optimizer was investigated. The results are presented in Figure 2 where the mean RSP relative error (top) and the respective standard deviation (bottom) are shown as a function of the number of projections and the particle type.

For all particles, a great reduction of the mean relative RSP error was achieved by using three projections instead of one, as shown in Figure 2. Using more than three projections did not further improve the results for the setup here investigated. The p-rad resulted in the largest mean relative errors of the RSP, followed by c-rad and α-rad.

### 3.3. Range uncertainties based on the charged particle type

Range uncertainties were evaluated using the method described in Section 2.4. For each charged particle type, the optimized calibration curves were obtained using three projections. The resulting calibration curves were then used to compute the WET of each tissues slab. The results are presented in Figure 3. For each considered physical thickness (from 1 cm to 30 cm), $10^4$ different tissue combinations were computed by sampling the PDF of the adult head phantom. In Figure 3, the bold lines represent the mean absolute WET-based range errors and the shaded area represents the respective standard deviation. For the simulated setup, the WET-based range errors were smaller for α-rad with maximum values of 1 mm.

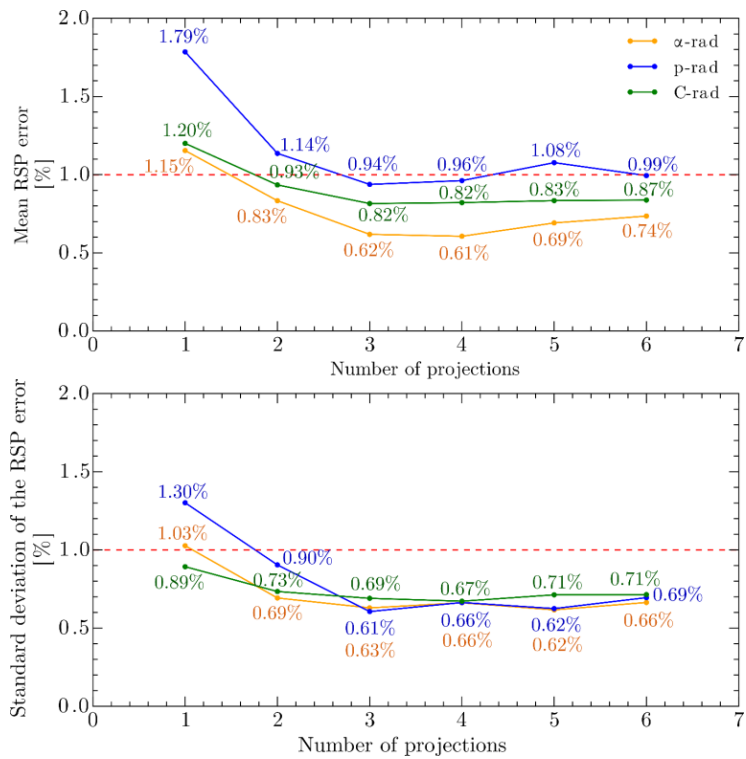

**Figure 2:** Top: mean RSP error for all optimized values as a function of the number of projections used. Bottom: Standard deviation of the RSP errors from the optimized curved as a function of the number of projections used.

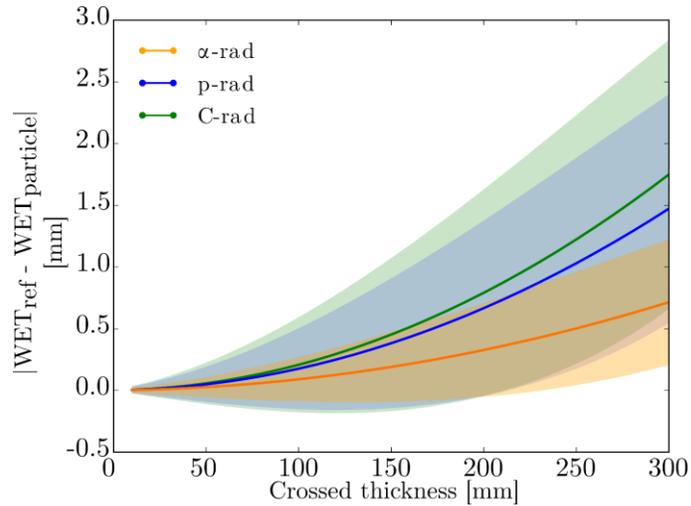

**Figure 3:** Absolute crossed WET errors. The results were computed using the optimized calibration curves with three projections for each considered charged particle. A sampling of 1 mm was considered for each crossed thickness. The WET was calculated using MC based method where for each thickness $10^4$ tissue combinations were considered. The MC calibration curve was used as reference.

### 3.4. Dose delivery based on the charged particle type

The dose delivered by each incident beam type, *i.e.*, proton, helium or carbon ions, was calculated for the presented setup. The dose delivered per projections was 1 µGy for p-rad, 8 µGy for α-rad and 40 µGy for c-rad.

## 4. Discussion

The aim of this study was to determine 1) the optimal charged particle type, 2) the optimal number of projections and 3) the best tissue segmentation method to obtain a patient-specific HU-RSP calibration curve.

Three tissue segmentation strategies were first proposed. The optimal tissue sampling method was identified to be the *k-means* clustering method (mean error of 0.62±0.63%). Both, the distortion and evenly selecting the tissues from the PDF, resulted in mean errors of the calibration curves larger than 1.5±0.7% (Figure 1(a)).

The optimal number of clusters to balance noise and accuracy was then determined. The results showed that selecting ≈ 1% of the HU values present in the adult head phantom was enough for an accurate optimization. A smaller set led to a larger mean relative error of the optimized curve. A larger set, on the other hand, resulted in larger fluctuations of the optimized RSP values. These results demonstrate the importance of proper tissue segmentation in the calibration procedure.

The effect of using multiple projections from different angles for the optimization was assessed for the different charged particles. Best results were found when using three projections (mean error of 0.62%±0.63 for α-rad) since the lowest standard deviation was achieved. No further improvement was seen when using more projections for the investigated setup.

The impact of using different charged particles to generate the radiographies was assessed. The largest mean RSP errors (Figure 2) were obtained for p-rad, independently of the number of projections. However, the metric of interest in comparing methods is the range uncertainty as it combines the RSPs errors with the frequency distribution of the tissues within the phantom. WET range errors depend on the beam direction, *i.e.*, which materials are being crossed, and therefore, a bias can be introduced by choosing a certain direction for the evaluation. By using the technique proposed by Bär *et al.* [21] this direction dependency was removed and a more general estimate of the range errors could be given. Lowest WET errors were achieved for the optimization performed with α-rad. These results are in-line with previous results from our group. Indeed, it has been found that the path estimate algorithm performs best when evaluating helium particles trajectories [22] compared to other charged particles (proton and carbon). Furthermore, α-rad suffers less from fluence-reducing nuclear reaction events compared to c-rad. Nuclear interactions increase the sparseness of the system matrix (A) and, hence, reduce the accuracy of the obtained calibration curve.

Increasing the initial statistics to retain a higher equal number of events for the optimization would increase the dose to the object. For the setup investigated here, using c-rad led to an increase of the dose by a factor 5 compared to α-rad.

The particularly high precision of the technique can be appreciated in Figure 3. These results demonstrate that for all considered particles in the shallow treatment area (< 5 cm), the errors in the calculation of the crossed WET were under 0.2 mm. For a depth greater than 20 cm, a considerable difference can be seen between helium and the other two particles. Using α-rad led to a mean error below 0.2 mm and a standard deviation below 0.5 mm, while for the other particles a mean error larger than 0.5 cm was observed. For comparison, the actual clinical accuracy is 3.5%, *i.e.*, for a range of 20 cm a maximal error of 7 mm can be expected. Of course, this comparison neglects the potential errors introduced by the experimental reality (*e.g.* detector uncertainties, beam line uncertainties) and serves only as a reference to what can be achieved by this technique.

The drawback of this method, independently of the particle type, is that it is limited by patient motion and anatomical changes, which can cause a mis-registration between the radiography and the X-ray CT. Further investigation that tackles this problem and an experimental validation of the proposed method are necessary.

## 5. Conclusion

This study aimed to investigate the optimal charged particle type and number of projections to determine a HU-RSP calibration curve to be used in charged particle therapy. Different strategies to segment and select the materials used to optimize the HU-RSP calibration curve were assessed. Only 1% of the HU values of the X-ray CT and one radiography were necessary for the optimization method to provide accurate results. The results were compared for three different charged particle radiography types: proton (p-rad), helium (α-rad) and carbon (c-rad). The results suggest that α-rad might be the method of choice for future charged particle treatment planning.

## Acknowledgments

This work has been funded by the grant SFRH/BD/85749/2012 from Fundação para a Ciência e a Tecnologia (FCT).